\documentclass{moriond}

\def\Journal#1#2#3#4{{#1} {\bf #2}, #3 (#4)}

\def\PRD{{\em Phys. Rev.} D}

\def\be{\begin{equation}}
\def\ee{\end{equation}}
\def\bea{\begin{eqnarray}}
\def\eea{\end{eqnarray}}

\newcommand{\Photo}{}
\newcommand{\chimera}{{\sc Chimera}}

\begin{document}
\vspace*{4cm}
\title{MULTIMESSENGERS FROM 3D CORE-COLLAPSE SUPERNOVAE}
 
\author{ K.~N.~YAKUNIN$^{ab}$, P.~MARRONETTI$^c$, A. MEZZACAPPA$^{ab}$, O.~E.~B.~MESSER$^{ade}$, E.~LENTZ$^{abe}$, S.~BRUENN$^f$, W.~RAPHAEL~HIX$^{ac}$, J.~A.~HARRIS$^{a}$}

\address{
$^a$Department of Physics and Astronomy, University of Tennessee, Knoxville, TN 37996 USA\\
$^b$Joint Institute for Computational Sciences, ORNL, Oak Ridge, TN 37831 USA\\
$^c$Physics Division, National Science Foundation, Arlington, VA 22230 USA\\
$^d$National Center for Computational Sciences, ORNL, Oak Ridge, TN 37831 USA\\
$^e$Physics Division, ORNL, Oak Ridge, TN 37831, USA\\
$^f$Department of Physics, Florida Atlantic University, Boca Raton, FL 33431, USA}

\maketitle\abstracts{
We present gravitational wave and neutrino signatures obtained in our \emph{ab initio} 3D core-collapse supernova simulation of a 15\,M$_\odot$ non-rotating progenitor with the {\chimera} code. Observations of neutrinos emitted by the forming neutron star and the gravitational waves produced by hydrodynamic instabilities are the most promising channel of direct information about the supernova engine. Both signals show different phases of the supernova evolution.}


The era of multimessenger astronomy is about to begin as an advanced generation of gravitational wave detectors will come on-line this year.
Core-Collapse Supernovae (CCSN) are among the most promising sources for multi-messenger astronomy due to strong electromagnetic and neutrino signals, as well as powerful gravitational wave (GW) bursts. 
Multimessenger observations could help resolve a number of open questions concerning the physics of CCSN such as: 1) What collapse mechanisms can we confirm or reject? 
2) Can GW detectors provide an early warning to EM observers?
3) What happens in CCSN before light and neutrinos break free?

\begin{figure}
\begin{minipage}{0.50\linewidth}
\centerline{\includegraphics[width=1.0\linewidth]{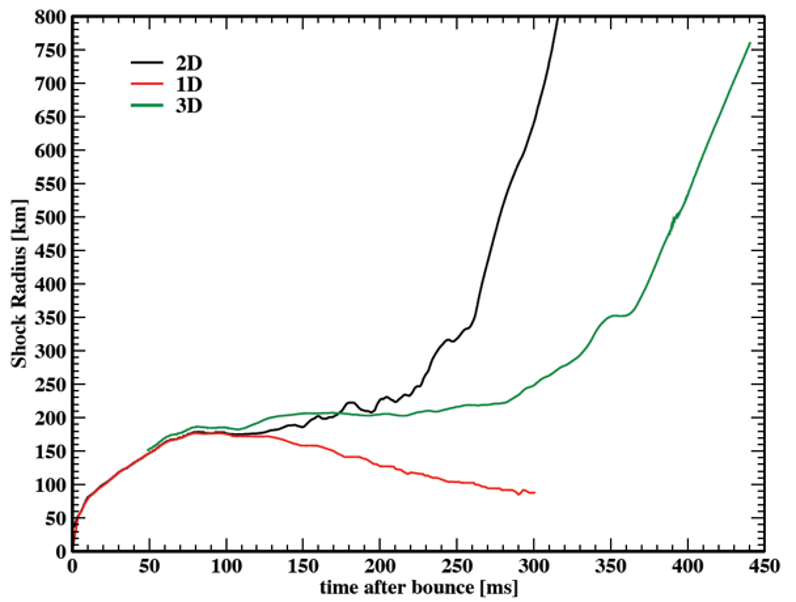}}
\end{minipage}
\hfill
\begin{minipage}{0.50\linewidth}
\centerline{\includegraphics[width=1.0\linewidth]{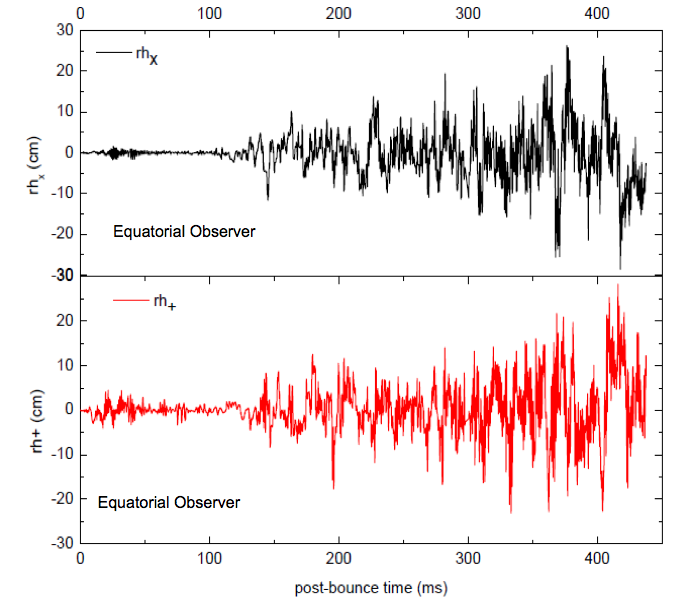}}
\end{minipage}
\caption[]{Evolution of the shock trajectory from our 1D model and the angle-averaged shock trajectories from our 2D and 3D models, all for the same 15M$_\odot$ progenitor (left). Gravitational wave polarizations $rh_+$  and $rh_\times$ as a function of post-bounce time seen by an observer on the equator (right).}
\label{fig:signals}
\end{figure}

  
In order to address these questions, we study the GW emission in a 3D model performed with the neutrino-hydrodynamics code \chimera~\cite{Bruenn14}, which is composed of five major modules: hydrodynamics, neutrino transport, self-gravity, a nuclear equation of state, and a nuclear reaction network. We evolve a non-rotating model corresponding to a zero-age main sequence progenitor of 15\,M$_\odot$~\cite{Woosley07}, on an adaptive spherical-polar mesh with resolution ~$512(r)\times180(\theta)\times180(\phi)$. 
This model was simulated using the Lattimer--Swesty equation
of state (EoS) with $K = 220$~MeV for $\rho > 10^{11} \textrm{g}\,\textrm{cm}^{-3}$, and an enhanced version of the Cooperstein EoS for $\rho < 10^{11} \textrm{g}\,\textrm{cm}^{-3}$. The simulation exhibits shock revival and the development of neutrino-driven explosions, a unique feature for first-principle simulations from progenitors with canonical CCSN masses (Fig.~\ref{fig:signals}~left). All the main phases of supernova dynamics can be seen in the gravitational waveforms (Fig.~\ref{fig:signals}~right): prompt convection, standing accretion shock instability (SASI), neutrino-driven convection, and formation of accretion downflows impinging on the surface of the proto-neutron star. The frequency of the gravitational wave signals tends to increase during the first 500 ms of post-bounce evolution.

\begin{figure}[h]
    \includegraphics[width=0.5\textwidth]{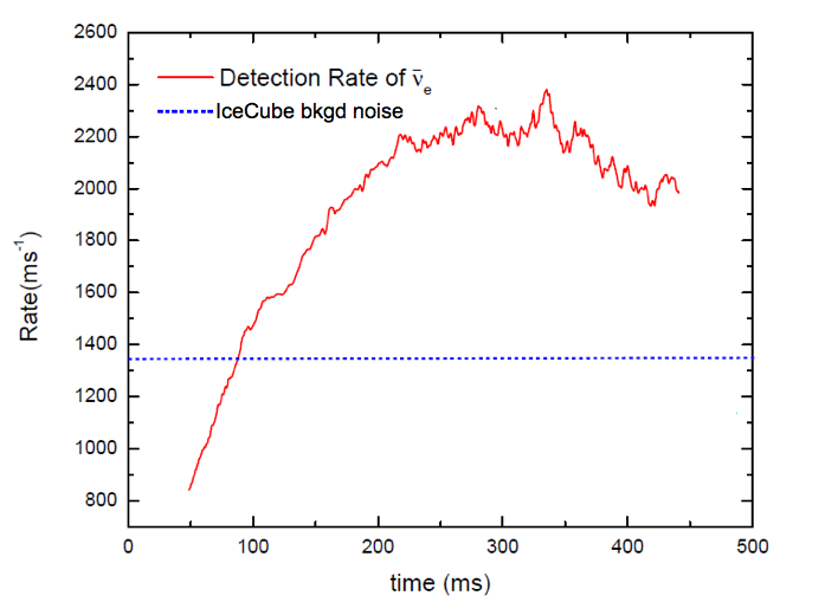}
    \caption{Detection rate of $\bar\nu_e$ in IceCube for Galactic CCSN.} 
\label{fig:icecube}
\end{figure}

Low-energy neutrinos (LENs) will be an important multi-messenger partner to GWs from CCSN. A CCSN produces 10--160 MeV neutrinos (all flavors) over a few  tens of seconds. The estimation of antineutrino rate detection in IceCube ~\cite{IceCube} presented in Fig.~\ref{fig:icecube} was done using Eq. (1) of Lund ~\emph{et al.}~\cite{Lund12}.


The SASI, with characteristic frequencies of 50--100 Hz, strongly imprints the neutrino signals observable by large Cherenkov detectors for Galactic CCSN.  If neutrino-driven convection dominates, the pre-explosion time variations of the neutrino flux are expected to exhibit smaller amplitude and higher frequency variations. Hence, the neutrino signal of the next Galactic CCSN may observationally constrain the contribution of neutrino-driven convection and SASI~\cite{Ott13}.\\

{\bf{Acknowledgments:}} This research was supported by the U.S. Department of Energy Offices of Nuclear Physics and Advanced Scientific Computing Research and the NASA Astrophysics Theory and Fundamental Physics Program (NNH11AQ72I). PM is supported by the National Science Foundation through its employee IR/D program. The opinions and conclusions expressed herein are those of the authors and do not represent the National Science Foundation.

\end{document}